\documentclass[aps,prl,twocolumn,superscriptaddress,showpacs,showkeys]{revtex4}
\usepackage{epsfig}
\usepackage{graphicx}
\usepackage{amsmath}
\usepackage[english]{babel}

\begin{document}

\title{Continuous percolation phase transitions of random networks under a generalized Achlioptas process}

\author{Jingfang Fan}
\affiliation{State Key Laboratory of Theoretical Physics, Institute of Theoretical Physics,Chinese Academy of Sciences, P.O. Box 2735, Beijing 100190, China}
\author{Maoxin Liu }
\affiliation{State Key Laboratory of Theoretical Physics, Institute of Theoretical Physics,Chinese Academy of Sciences, P.O. Box 2735, Beijing 100190, China}
\author{Liangsheng Li}
\affiliation{Key Laboratory of Cluster Science of Ministry of Education and Department of Physics, Beijing Institute of Technology, Beijing 100081, China}
\author{Xiaosong Chen}
\affiliation{State Key Laboratory of Theoretical Physics, Institute of Theoretical Physics,Chinese Academy of Sciences, P.O. Box 2735, Beijing 100190, China}
\email{chenxs@itp.ac.cn}
\date{\today}
\begin{abstract}

Using the finite-size scaling, we have investigated the percolation phase transitions of evolving random networks under a generalized Achlioptas process (GAP). During this GAP, the edge with minimum product of two connecting cluster sizes is taken with a probability $p$ from two randomly chosen edges. This model becomes the Erd\H os-R\'enyi network at $p=0.5$ and the random network under the Achlioptas process at $p=1$. Using both the fixed point of $s_2/s_1$ and the straight line of $\ln s_1$, where $s_1$ and $s_2$ are the reduced sizes of the largest and the second largest cluster, we demonstrate that the phase transitions of this model are continuous for $0.5 \le p \le 1$. From the slopes of $\ln s_1$ and $\ln (s_2/s_1)'$ at the critical point we get the critical exponents $\beta$ and $\nu$, which depend on $p$. Therefore the universality class of this model should be characterized by $p$ also.

\end{abstract}

\maketitle

The modern theory of complex networks \cite{Albert,Watts,Newman} has opened new perspectives in the study of complex systems in nature, social and economic systems, technical infrastructures and many other fields. The macroscopic properties of complex networks emerge from the interactions among individual constituents. The percolation phase transition of complex networks is an interesting macroscopic property and can be studied with the percolation theory in statistical physics \cite{staufferbook}. It has been pointed out that the random network undergoes a continuous percolation phase transition during the random process \cite{bollobas}. The critical phenomena of complex networks are reviewed in Ref.\cite{dorogovtsev08}.

However, Achlioptas et al.\cite{achlioptas09} reported that the percolation phase transition of random networks becomes discontinuous under the Achlioptas process (AP). During this AP, two unoccupied edges are chosen randomly and the edge with minimum product of the connecting cluster sizes is taken as the next occupied bond. This discontinuous phase transition in networks is called the explosive percolation \cite{achlioptas09}. The AP was applied later to other networks \cite{ziff09,ChoP09,Rad09} and some similar rules were also introduced \cite{ChoP09,Rad09,FriPRL09,ChoPRE10-1,SouzaPRL10,AraujoPRL10,MannaPhysicaA10}. Using the finite-size scaling, the explosive percolation phase transition was also investigated \cite{ChoPRE10-2,ZiffPRE10}. These works support that the explosive percolation is discontinuous. Recently, Costa et al.\cite{CostaPRL10} showed that the explosive percolation transition is actually a continuous phase transition with a uniquely small critical exponent.
Riordan and Warnke \cite{Riordan} prove mathematically that all Achlioptas processes have continuous phase transitions.

In this letter we investigate the percolation phase transitions of random networks under a generalized Achlioptas process (GAP). In the GAP, two unoccupied edges are chosen randomly and the edge with minimum product of two connecting cluster sizes is taken with a probability $p$ \cite{Liu}. At $p=0.5$, this model becomes the Erd\"os-R\'enyi (ER) network \cite{ER}. At $p=1$, the GAP is equivalent to the Achlioptas process which suppress the appearance of larger clusters. With the increase of $p$ from $0.5$ to $1$, this suppression is switched on gradually. After investigating the percolation phase transitions at different $p$, we can get more understanding about these phase transitions and make a more reliable conclusion about the character of the percolation phase transition at $p=1$.

In our Monte carlo simulations, we begin with $N$ isolated nodes and then connect them with the edges added through the GAP. The network obtained is characterized by $N$, the number of edges $N_r$ and the probability parameter $p$ of GAP. For $p <1$, we take $N=(4,~8,~16,~32,~64,~128)\times 10^4$ and run $500,000$ steps in each simulation. At $p=1$, we choose much larger system sizes $N=(512,~1024,~2048)\times 10^4$ to reach the asymptotic region of finite-size effects and run $100,000$ steps in each simulation.

\begin{figure}
  \begin{center}
 \includegraphics[width=.5\textwidth]{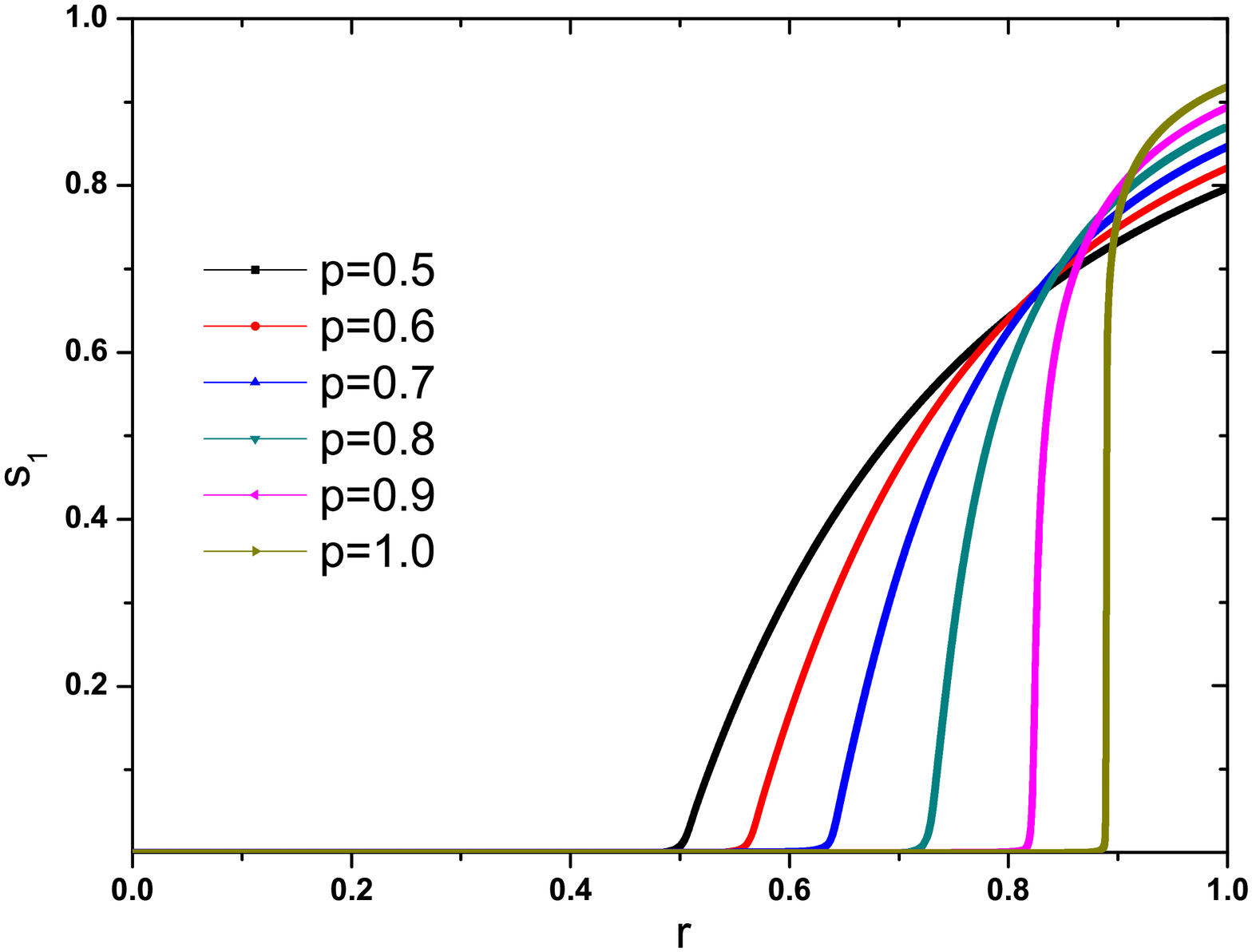}
  \end{center}
  \caption{ Reduced size of the largest cluster at $N=1.28 \times 10^6$.}
\label{fig1}
\end{figure}

For the cluster ranked $R$ and with size $S_R (N_r,~N;~p)$, we define its reduced size as
\begin{equation}
s_{R}(r,~N;~p) \equiv S_{R}(N_r,~N;~p)/N\;,
\label{eq:perc_str1}
\end{equation}
where $r\equiv N_r/N$ is the reduced edge number. The reduced size of the largest cluster $s_{1}(r,N;p)$ at $N=1.28\times 10^6$ is shown in Fig.~\ref{fig1}. $s_1$ becomes finite for $r > r_c$ and there is a percolation phase transition. The transition point $r_c$ increases with $p$ and the corresponding phase transition becomes sharper.

If the percolation phase transition is continuous, the reduced sizes $s_R (r,~N;p)$ should follow a finite-size scaling form \cite{privman1,privman2}
\begin{equation}
s_R(r,N;p) = N^{-\beta/\nu}\widetilde{s}_{R}(tN^{1/\nu};p),
\label{eq:perc_str2}
\end{equation}
where $t=(r-r_c)/r_c$  and $\nu$  is the critical exponent of the correlation length $\xi = \xi_{0}\left|t\right|^{-\nu} $. The scaling form in Eq. (\ref{eq:perc_str2}) is valid in the asymptotic critical region with $|t|\ll 1$ and $N \gg 1$.  Outside this region, additional correction terms should be taken into account.

From Eq. (\ref{eq:perc_str2}), we can obtain
\begin{equation}
s_2/s_1 = \widetilde{s}_2 (tN^{1/\nu};p)/ \widetilde{s}_1 (tN^{1/\nu};p) \equiv U(tN^{1/\nu};p)
\label{eq:perc_str7}
\end{equation}
and
\begin{equation}
\ln s_1(r,N;p)= -\beta/{\nu}\ln N+ \ln \widetilde{s}_{1}(tN^{1/\nu};p).
\label{eq:perc_str9}
\end{equation}
At a critical point, $s_2/s_1|_{r=r_c} = U(0;p)$ is a fixed point versus $N$ and $\ln s_{1}(r_c,N;p)=-\beta/{\nu}\ln N+ \ln \widetilde{s}_{1}(0;p)$ is a straight line versus $\ln N$. Using these properties, the critical point of complex network can be determined both from the fixed point of  $s_2/s_1$ and the straight line of $\ln s_1$ versus $\ln N $. The critical exponent ratio $\beta/\nu$ can be obtained from the slope of $\ln s_1$ at $r_c$.

\begin{figure}
\includegraphics[width=0.5\textwidth]{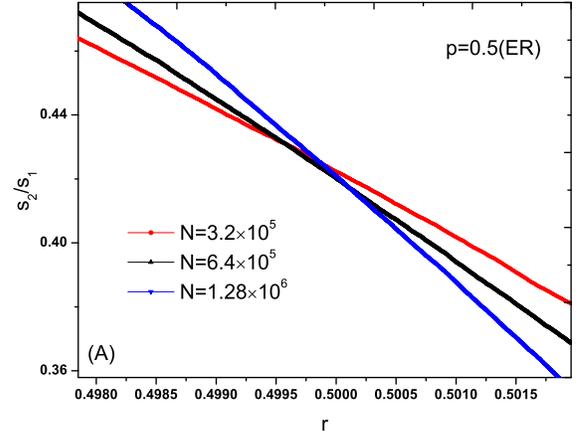}
\includegraphics[width=0.5\textwidth]{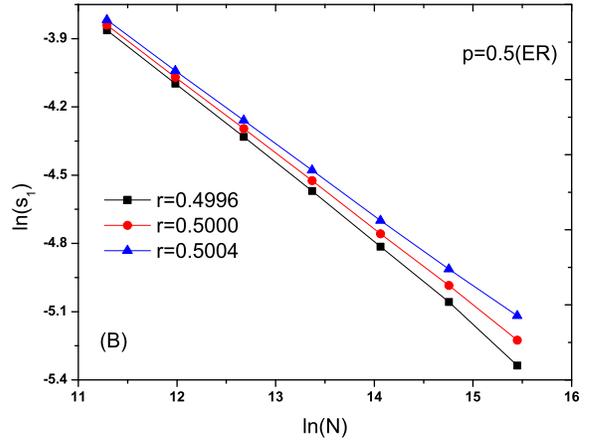}
\caption{(A) $s_2/s_1$ has a fixed point at $r^1_{c} = 0.5000\pm 0.0004$; (B) Log-log plot of the reduced size $s_{1}$ versus $N$. Around $r^2_c = 0.5000$,
the curvature of $\ln s_1$ changes its sign.
}
\label{fig2}
\end{figure}

 For $p = 0.5$, our model is equivalent to the ER model \cite{ER}.  Its continuous percolation phase transition can be studied with $s_2/s_1$ and $\ln s_1$. In Fig.~\ref{fig2}(A), $s_2/s_1$ is shown as a function of $r$ at different $N$. The different curves of $s_2/s_1$ for different $N$ have a fixed point at $r^1_c = 0.5000(4)$. In Fig.~\ref{fig2}(B), $\ln s_1$ is shown versus $\ln N$ for different $r$. The curvature of $\ln s_1$ is negative at $r = 0.4996$ and becomes positive at $r = 0.5004$. At $r = 0.5$, $\ln s_1$  is a straight line with slope $-0.33(1)$. We get $r^2_c=0.5000(4)$ in consistence with $r^1_c$. Our results of $r_c$ and $\beta/\nu$ agree with the exact results $r_c =0.5$ and $\beta/\nu = 1/3$ of the ER model \cite{ER}.

\begin{figure}
\includegraphics[width=.5\textwidth]{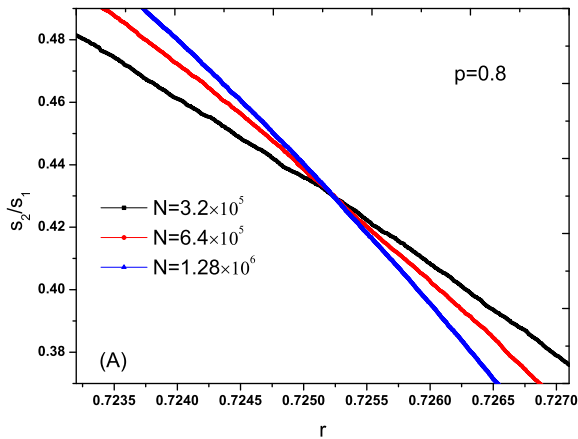}
\includegraphics[width=0.5\textwidth]{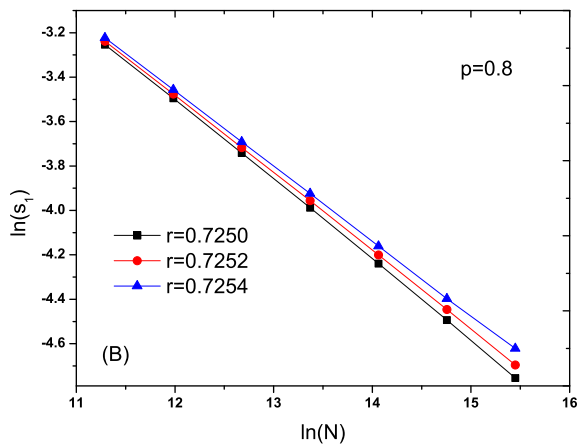}
\caption{(A)$s_2/s_1$ has a fixed point at $r^1_c = 0.7252\pm 0.0001$; (B)Log-log plot of the reduced size $s_1$ versus $N$ at different $r$. Around $r^2_c = 0.7252$, the curvature of $\ln s_1$ changes its sign.
}
\label{fig3}
\end{figure}

In Fig.~\ref{fig3},  we show $s_2/s_1$ and $\ln s_1$ of random network under the GAP at $p = 0.8$. From the fixed point of $s_2/s_1$, we get the critical point $ r^1_{c}=0.7252\pm 0.0001$. The curvature of $\ln s_1$ is negative at $r=0.7250$ and becomes positive at $r=0.7254$. So $r^2_c = 0.7252\pm 0.0002$, which is consistent with $r^1_c$.

\begin{figure}
\includegraphics[width=.5\textwidth]{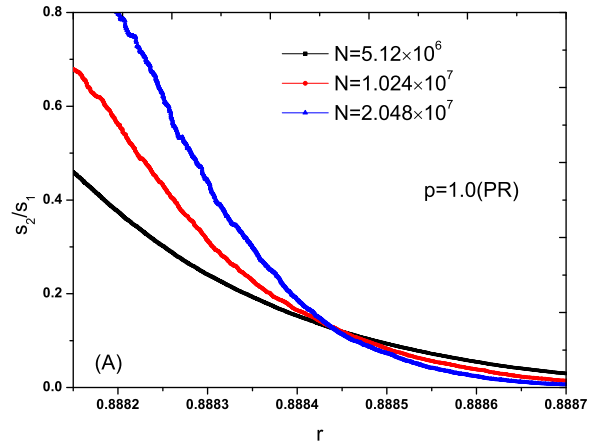}
\includegraphics[width=.5\textwidth]{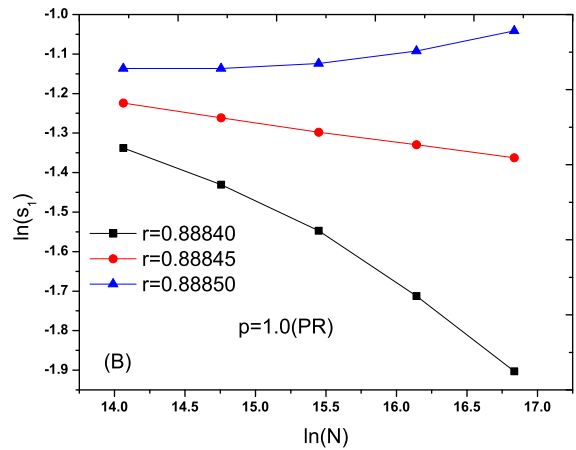}
\caption{(A) $s_2/s_1$ has a fixed point at $r^1_c = 0.88844\pm0.00002$.
(B) Log-log plot of the reduced size $s_{1}$ versus $N$. At $r^2_c = 0.88845$, the slope of $\ln s_1$ versus $\ln N$ gives $\beta/\nu \approx 0.04$.
 }
\label{fig4}
\end{figure}

 For $p = 1.0$, our model becomes the PR model of Ref.\cite{achlioptas09}. $s_{2}/s_{1}$ at different $N$ are shown in Fig.~\ref{fig4}(A) and there is a fixed-point at $r^1_c=0.88844(2)$. The curvature of $\ln s_{1}$ is negative at $r = 0.88840$ and becomes positive at $r=0.88850$. We obtain $r^2_c=0.88845(5)$ consistent with $r^1_c$. From the slope of $\ln s_1$ at $r^2_c$, we get $\beta/\nu = 0.04(1)$. Our $r_c$ and $\beta/\nu$ at $p=1$ agree well with Ref.\cite{Grassberger}.

To determine the critical exponent $\nu$ of correlation length, we study the derivative
\begin{equation}
(s_2/s_1)' \equiv \frac{\partial(s_2/s_1)}{\partial r} = N^{1/\nu}r^{-1}_cU'(tN^{1/\nu};p)\;.
\label{eq:perc_str11}
\end{equation}
At a critical point, we have
\begin{equation}
\ln (s_2/s_1)'|_{r=r_c} = (1/\nu)\ln N + ln U'(0;p) - ln r_{c},
\label{eq:perc_str11}
\end{equation}
which is a straight line with slope $1/\nu$. From our Monte Carlo data, we can calculate $1/\nu$ at different $p$, which are summarized in Table I.

\begin{table}[htbp]
\centering  %
\caption{\label{comparison}Critical reduced edge number $r_{c}$, critical exponent ratio $\beta/\nu$ and inverse of critical exponent $1/\nu$. We obtain $r^{1}_{c}$ from $s_{2}/s_{1}$ and $r^{2}_{c}$ from $\ln s_{1}$.}
\begin{tabular}{c|c|c|c|c}
\hline
p  &$r^{1}_{c}$  &$r^{2}_{c}$   &$\beta/\nu$  &$1/\nu$\\ \hline
0.5  &0.5000(4) &0.5000(4) &0.33(1) &0.33(1)\\         %
0.6  &0.5599(1) &0.5599(3) &0.32(1) &0.33(1)\\        %
0.7 &0.6349(2) &0.6349(3) &0.32(1) &0.34(1)\\
0.8 &0.7252(1) &0.7252(2) &0.33(1) &0.35(1)\\
0.9 &0.8216(1) &0.8216(3) &0.31(1) &0.38(1)\\
0.95 &0.8603(2) &0.8604(2) &0.15(1) &0.46(1)\\
1.0 & 0.88844(2) &0.88845(5) &0.04(1) &0.50(1) \\ \hline
\end{tabular}
\end{table}

\begin{figure}
\includegraphics[width=.5\textwidth]{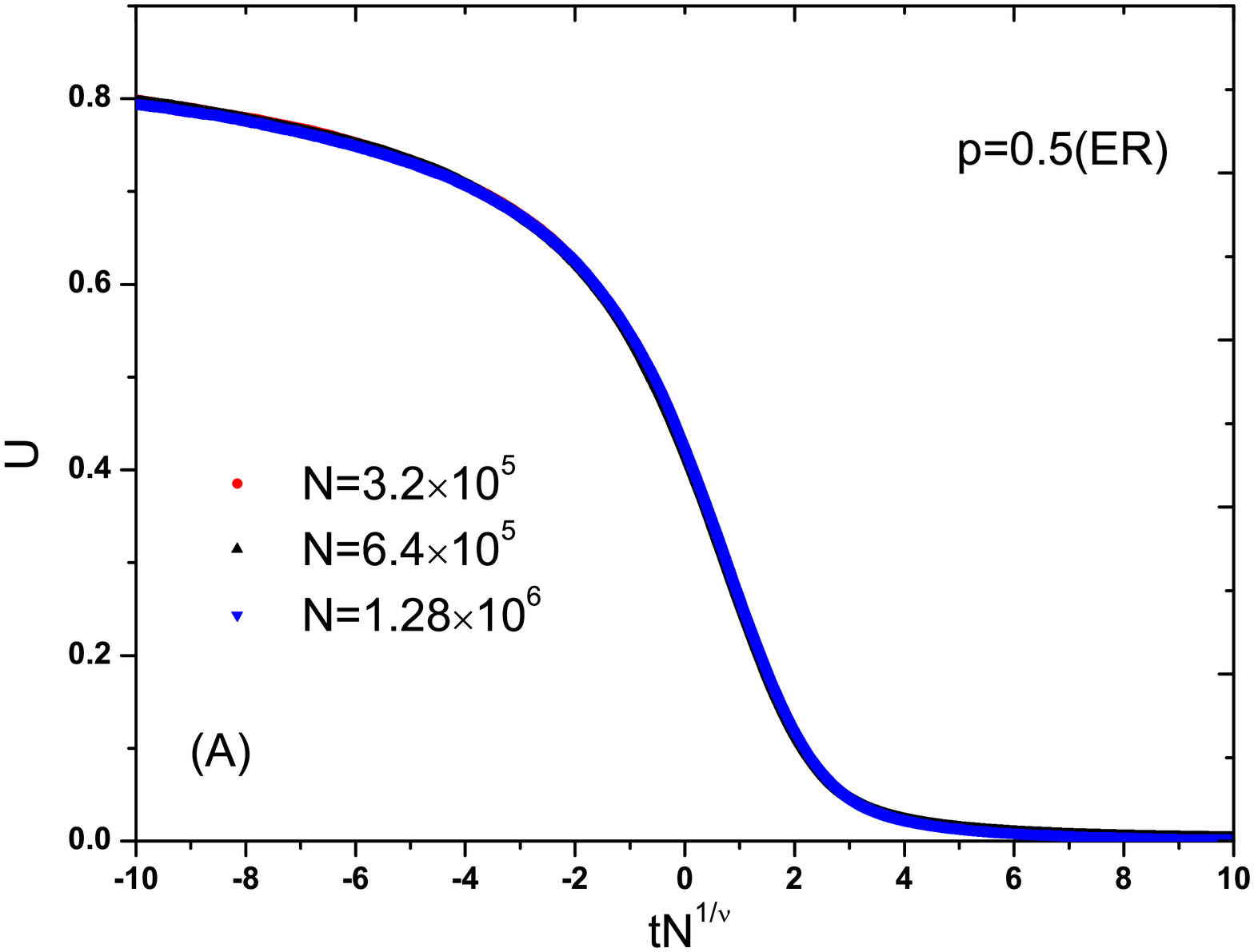}
\includegraphics[width=.5\textwidth]{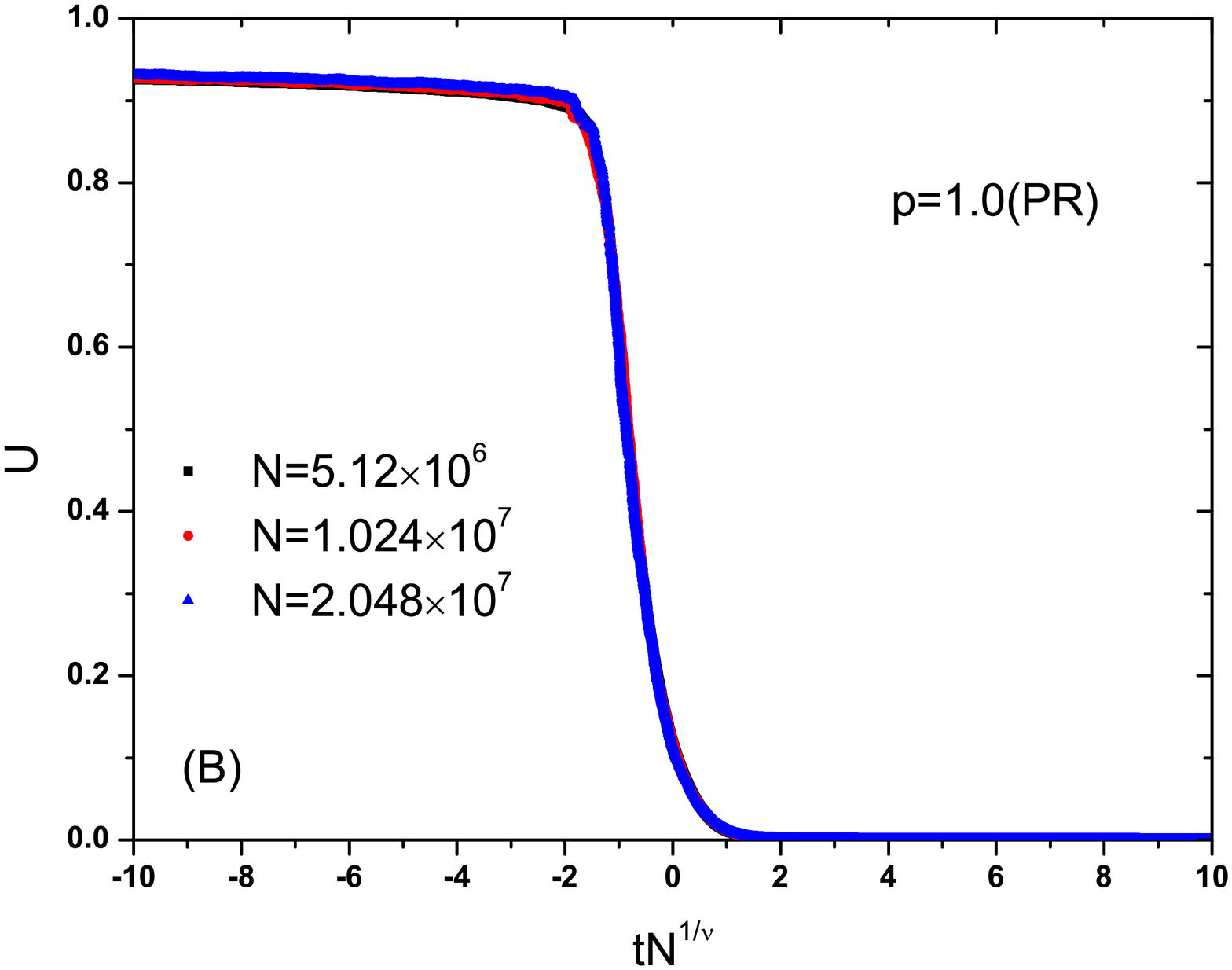}
\caption{Finite-size scaling function $U(tN^{1/\nu};p)$ at
(A)  $p = 0.5$ with $1/\nu = 0.33$ and (B) $p = 1.0$ with $1/\nu = 0.5$.
}
\label{fig6}
\end{figure}

\begin{figure}
\includegraphics[width=.5\textwidth]{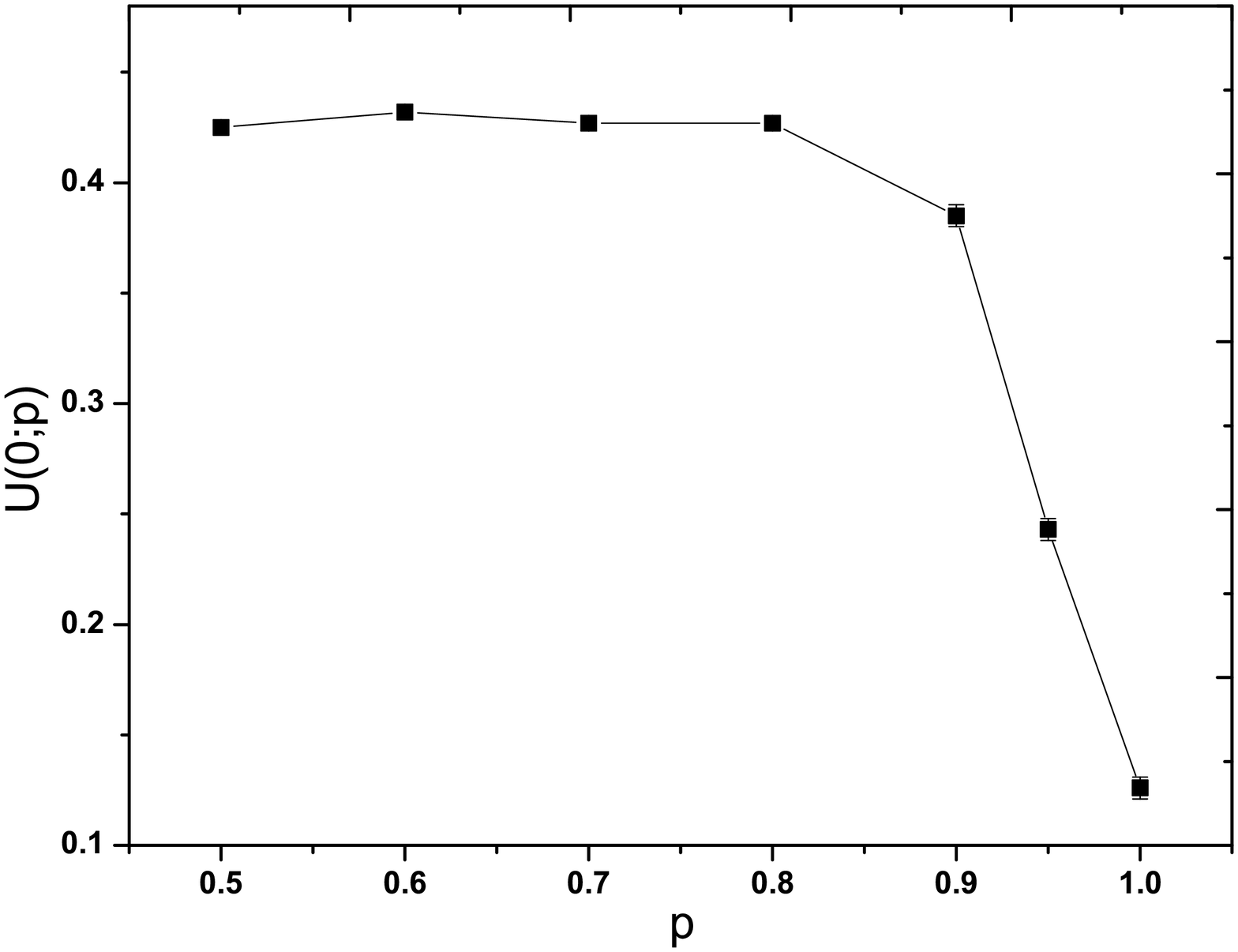}
\caption{$U(0;p)$ with error bars smaller than the symbol.
}
\label{fig8}
\end{figure}

Using $r_c$ and $1/\nu$ given in Table I , we can define a finite-size scaling variable $tN^{1/\nu}$ at each $p$. With this scaling variable, our Monte Carlo simulation data of $s_2/s_1(r,N;p)$ at a given $p$ and different $N$ collapse into a finite-size scaling function $U(tN^{1/\nu};p)$, which are shown in Fig.~\ref{fig6} for $p=0.5$ and $p=1$. The $p$-dependence of $U(0;p)$ is demonstrated in Fig.~\ref{fig8}.

In summary, we have studied the percolation phase transitions of evolving random networks under a generalized Achlioptas process that the edge with minimum product of two connecting cluster sizes is taken with a probability $p$ from two randomly chosen edges. This model becomes the ER network \cite{ER} at $p=0.5$ and the random network under the Achlioptas process \cite{achlioptas09} at $p=1$. Using the finite-size scaling, the percolation phase transitions of this model are studied. From the finite-size scaling forms of the largest cluster size $s_1(r,N;p)$ and the second largest cluster size $s_2(r,N;p)$, the fixed point of $s_2/s_1$ and the straight line of $\ln s_1$ versus $\ln N$ can be used to determine the critical points. It has been found that the critical points $r_c$ determined from $s_2/s_1$ and $\ln s_1$ are consistent and increase with $p$. From the slopes of $\ln s_1$ and $\ln (s_2/s_1)'$ at $r_c$, we can obtain the critical exponent ratio $\beta/\nu$ and $1/\nu$ respectively. $\beta/\nu$ decreases from $0.33(1)$ at $p=0.5$ to $0.04(1)$ at $p=1$. $1/\nu$ increases from $0.33(1)$ at $p=0.5$ to $0.50(1)$ $p=1$. With the scaling variable $tN^{1/\nu}$, the Monte Carlo data of $s_2/s_1(r,N;p)$ at a given $p$ and different $N$ collapse into a scaling function $U(tN^{1/\nu};p)$. The percolation phase transitions of random networks under the GAP are continuous always and its universality class depends on the probability parameter $p$.

This work is supported by the National Natural Science Foundation
of China under grant 10835005.

\end{document}